\documentclass[aps,twocolumn]{revtex4}
\usepackage{graphicx}			
\input psfig.sty			

\def\wisk#1{\ifmmode{#1}\else{$#1$}\fi}

\def\deg    {\wisk{^\circ}}

\begin{document}
\normalsize

\title{Design and Calibration of a Cryogenic Blackbody Calibrator
at Centimeter Wavelengths}

\author{A. Kogut}
\email{Alan.J.Kogut@nasa.gov}
\affiliation{NASA/Goddard Space Flight Center, 
Laboratory for Astronomy and Solar Physics,
Greenbelt, Maryland 20771}

\author{E. Wollack}
\affiliation{NASA/Goddard Space Flight Center, 
Laboratory for Astronomy and Solar Physics,
Greenbelt, Maryland 20771}

\author{D. J. Fixsen}
\affiliation{~SSAI Code 685, 
NASA/GSFC Laboratory for Astronomy and Solar Physics,
Greenbelt, Maryland 20771}

\author{M. Limon}
\affiliation{~SSAI Code 685, 
NASA/GSFC Laboratory for Astronomy and Solar Physics,
Greenbelt, Maryland 20771}

\author{P. Mirel}
\affiliation{~SSAI Code 685, 
NASA/GSFC Laboratory for Astronomy and Solar Physics,
Greenbelt, Maryland 20771}

\author{S. Levin}
\affiliation{Jet Propulsion Laboratory,
California Institute of Technology,
4800 Oak Grove Drive,
Pasadena, California 91109 }

\author{M. Seiffert}
\affiliation{Jet Propulsion Laboratory,
California Institute of Technology,
4800 Oak Grove Drive,
Pasadena, California 91109 }

\author{P. M. Lubin}
\affiliation{Department of Physics,
University of California at Santa Barbara,
Santa Barbara, California, 93106}


\begin{abstract}
We describe the design and calibration
of an external cryogenic blackbody calibrator 
used for the first two flights of
the Absolute Radiometer for Cosmology, Astrophysics, 
and Diffuse Emission (ARCADE) instrument.
The calibrator consists of a microwave absorber
weakly coupled to a superfluid liquid helium bath.
Half-wave corrugations viewed 30\deg ~off axis
reduce the return loss below -35 dB.
Ruthenium oxide resistive thermometers
embedded within the absorber
monitor the temperature
across the face of the calibrator.
The thermal calibration 
transfers the calibration of a reference thermometer
to the flight thermometers
using the flight thermometer readout system.
Data taken near the superfluid transition
in 8 independent calibrations 4 years apart
agree within 0.3 mK,
providing an independent verification
of the thermometer calibration
at temperatures near that of the cosmic microwave background.
\end{abstract}

\pacs{}
\keywords{}
\maketitle

\section{INTRODUCTION}
The Absolute Radiometer for Cosmology, Astrophysics, and Diffuse Emission
(ARCADE) is a balloon-borne instrument
designed to measure the temperature of the cosmic microwave
background at centimeter wavelengths
\cite{kogut/etal:2004}.
ARCADE uses a set of narrow-band cryogenic radiometers
to compare the sky
to an external blackbody calibration target,
in order to detect or limit deviations from a blackbody spectrum.
At centimeter wavelengths,
raw sensitivity is not an important design criterion;
the instrument is designed instead
to reduce or eliminate
major sources of systematic uncertainty.
The instrument is fully cryogenic;
all major components are independently temperature-controlled 
to remain near 2.7 K, isothermal with the signal from deep space.
Boiloff helium vapor,
vented through the aperture plane,
forms a barrier between the instrument and the atmosphere
at 35 km altitude;
there are no windows or warm optics to correct.
An independently controlled blackbody calibrator 
located on the antenna aperture plane
rotates to cover each antenna in turn,
so that each antenna alternately views the sky 
or a known blackbody.
The cryogenic design minimizes the effect of internal reflection
or absorption:
any residual instrumental signals cancel to first order
in the sky--calibrator comparison.
All radiometers view the same calibrator in turn,
eliminating cross-calibration uncertainties
when comparing the sky temperature between different frequency channels.

The external calibrator
is critical to the ARCADE experiment.
It establishes a common blackbody reference
for all frequency channels,
and establishes an absolute temperature reference
for comparison with other instruments.
Assuring the electromagnetic performance of the calibrator,
though non-trivial, is straightforward.
The ability to achieve ARCADE's science goals 
can thus be traced to precision and accuracy
with which the physical temperature distribution
within the calibrator can be monitored.

\section{Calibrator Design}
The experimental design places multiple constraints
on the calibrator.
It must maintain stable cryogenic temperatures
while located at the mouth of an open liquid helium bucket Dewar,
and must be compact enough
to avoid spillover from the sidelobes of antennas viewing the sky.
The absorber should be nearly isothermal,
so that a limited number of temperature sensors
can sample the temperature distribution across the calibrator
with sufficient precision to provide the necessary radiometric calibration.

\begin{figure}[t]
\includegraphics[width=3.25in]{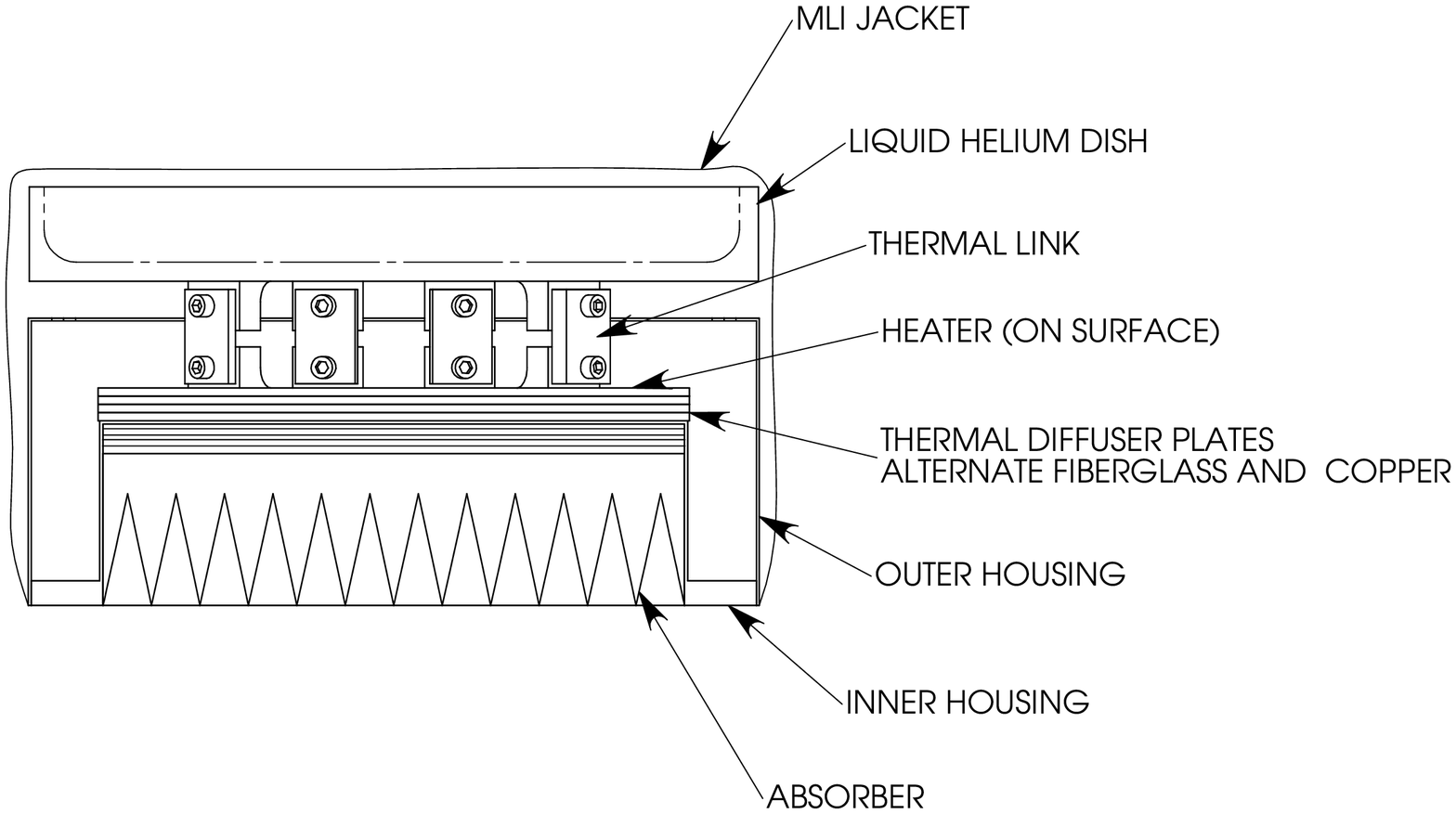}
\caption{Schematic of ARCADE external blackbody calibrator.
The microwave absorber (Eccosorb CR-112)
is weakly coupled to a 
superfluid liquid helium reservoir
through a series of copper/fiberglass buffer plates.
The absorber is 21 cm across,
with corrugated front surface
to reduce reflected power.
\label{target_schematic} }
\end{figure}

Figure \ref{target_schematic}
shows the schematic design
of the calibration target.
It consists of a microwave absorber
(Emerson \& Cuming Eccosorb CR-112, an iron-loaded epoxy)
cast with grooves in the front surface 
to reduce surface reflections.
To increase viscosity and reduce settling 
of the iron powder during the curing process,
we added 0.5\% cabosil powder by mass
to the Eccosorb prior to casting.
Absorber samples prepared at GSFC including the cabosil filler
have complex permittivity  
$e_r^\prime = 3.9$ and 
$e_r^{\prime\prime} = 0.5$,
for a loss tangent $\tan \delta = 0.13$ measured at 30 GHz.
The Eccosorb is mounted on a set of two thermally conductive 
oxygen-free copper plates 
separated by fiberglass thermal insulators,
each 1.6 mm in thickness.
Thermal control is achieved by heating the outermost copper plate,
which is in weak thermal contact with a superfluid helium reservoir.
Radial thermal gradients at each stage 
are reduced by the ratio of the thermal conductance of the buffer plates.
A two-stage design is sufficient
to reduce the thermal ``footprint'' of the heater resistors
well below 1 mK at the Eccosorb/copper interface.

The calibrator is located at the mouth of an open bucket Dewar
and must maintain temperatures near 2.7 K
despite heating from
the thermal control loop,
infrared emission from the atmosphere,
and residual condensation of atmospheric nitrogen.
A stainless steel shell protects the calibrator sides and top,
with a 20-layer reflective blanket providing additional
isolation from the environment.
Fountain-effect pumps transport superfluid liquid helium
from the main Dewar 
to a separate reservoir inside the calibrator,
providing the necessary cooling without requiring moving parts.
Boiloff gas from the calibrator reservoir vents through holes
in the top cover plate.

Eccosorb CR-112 has a high index of refraction ($n \approx 4$),
producing a power reflection coefficient
$R > 0.25$ at normal incidence
for frequencies below 30 GHz
\cite{hemmati/etal:1985}.
At millimeter wavelengths,
blackbody calibrators typically minimize reflections
using deep conical designs
to ensure multiple reflections
\cite{mather/etal:1999,gush/etal:1990}.
At longer wavelengths, this becomes impractical
due to size constraints.
The front Eccosorb surface is cast instead
into corrugated grooves.
The ARCADE antennas
view the calibrator at an angle $\theta = 30\deg$ from normal incidence
\cite{kogut/etal:2004}.
In the limit of a monochromatic incident plane wave,
the spacing of the corrugations
can be chosen so that reflections from one corrugation
are exactly out of phase with the neighboring corrugations,
cancelling the $m=0$ ``specular'' reflection from the calibrator
for all odd multiples of the incident wavelength.
Although the electric field at the antenna aperture
is not a plane wave,
the analysis provides some guidance to minimize reflections
within a compact geometry.
Yokimori \cite{yokimori:1984}
has examined the effect of changing the pitch/height ratio
for corrugated dielectric surfaces
and finds minimal reflection
for corrugation height 1--2 times the pitch.
The ARCADE calibrator uses corrugations 3 cm deep 
spaced 1.5 cm apart,
minimizing reflections from the Eccosorb surface
in both the 10 and 30 GHz channels.
An additional 1.3 cm of absorber between the bottom of the corrugations
and the copper mounting structure
provides additional absorption
to reduce reflections from the metal backing.

\begin{figure}[b]
\includegraphics[width=3.25in]{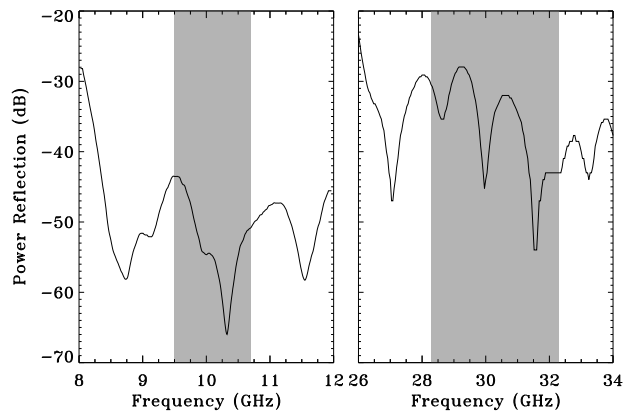}
\caption{Power reflection coefficient measured
using ARCADE calibrator and antennas
in the flight off-axis configuration.
The gray band shows the passband at each frequency range.
The gain-averaged in-band reflection is below -35 dB in each band.
\label{target_ref} }
\end{figure}

\section{MICROWAVE PERFORMANCE}
Calibration by an external target
corrects reflection and attenuation 
within the instrument to first order,
with the notable exception of the external calibrator reflectivity.
Reflections from the calibrator create a systematic offset in the
sky-calibrator comparison
proportional to the calibrator's power reflection coefficient 
and the temperature difference between the calibrator and the sky.
During flight, the calibrator will be commanded to different temperatures
that could vary from the sky temperature by as much as several K.
Achieving  1 mK or better accuracy for the sky temperature
thus requires calibrator power reflection below -30 dB.

\begin{figure}[t]
\includegraphics[width=3.25in]{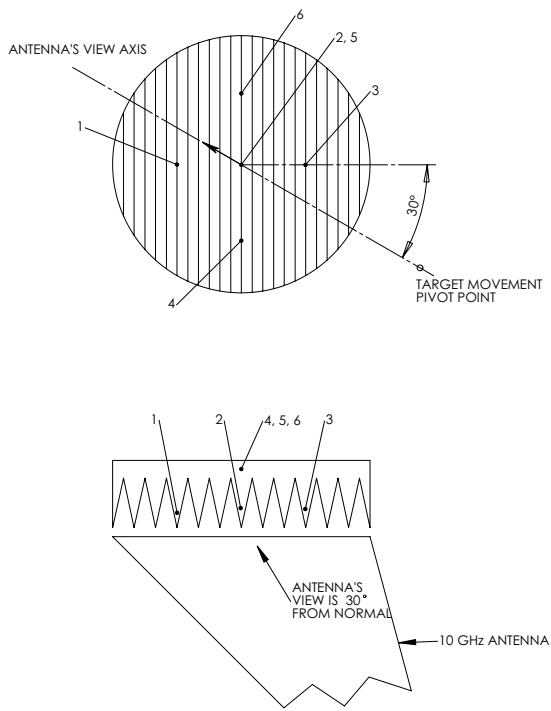}
\caption{Location of thermometers within the calibrator.
\label{therm_loc} }
\end{figure}

Figure \ref{target_ref} shows the power reflection coefficient
measured for both ARCADE frequency bands
using gated time-of-flight techniques at room temperature
with the antenna viewing the calibrator
in flight configuration (30\deg ~from normal incidence).
The maximum in-band reflection is -43 dB at 10 GHz,
with gain-weighted average -49 dB across the band.
At 30 GHz the maximum reflection is -28 dB,
with gain-weighted average -35 dB.
The results do not depend significantly
on the orientation of the absorber corrugations to the antenna.
Measurements of the antenna/calibrator coupling
using a vector network analyzer
provide similar limits $R < -35$ dB at 30 GHz.
As the absorber cools,
the epoxy filler shrinks
and the conductivity of the iron will increase,
increasing the reflectivity
and decreasing the absorption.
Measurements suggest an increase in power reflection
of 30 to 60 percent
at cryogenic temperatures
\cite{hemmati/etal:1985}.
Both bands still easily meet the -30 dB requirement.

\section{THERMOMETRY}
We monitor the calibrator temperature using
ruthenium oxide resistance thermometers,
chosen for their small size,
stability,
and large resistance change at cryogenic temperatures.
The thermometers consist of commercially available
thick-film chip resistors
(Dale Electronic RCWP-550)
with 76 $\mu$m Manganin leads
providing four-wire resistance measurements.
The resistance at room temperature is 10 k$\Omega$,
rising to 25 k$\Omega$ 
with slope $dT/dR = -0.158$ mK/$\Omega$
at 2.7 K.
Five thermometers embedded within the microwave absorber
monitor the temperature distribution,
while an additional two thermometers
mounted next to the heater on the outermost copper plate
provide feedback for thermal control.
Figure \ref{therm_loc} shows the location of the thermometers
in the calibrator.
Three thermometers
(labeled T1, T2, and T3)
are embedded near the tips of the Eccosorb corrugations,
with two other thermometers (T4 and T5)
located in the absorber
midway between the bottom of the corrugations
and the copper mounting plate.
Thermometers T2 and T5 are located atop each other
at the center of the calibrator 
to provide an estimate of vertical gradients
within the absorber.
After fabrication,
the entire calibrator,
including the embedded thermometers,
was cycled ten times 
between 300 K and 77 K
to relieve thermal stress.
A sixth thermometer embedded within the absorber
failed open
during this initial thermal cycling.
The calibrator has since been cooled
repeatedly below 2K with no adverse mechanical 
or electrical effects.

\begin{figure}[b]
\includegraphics[width=3.25in]{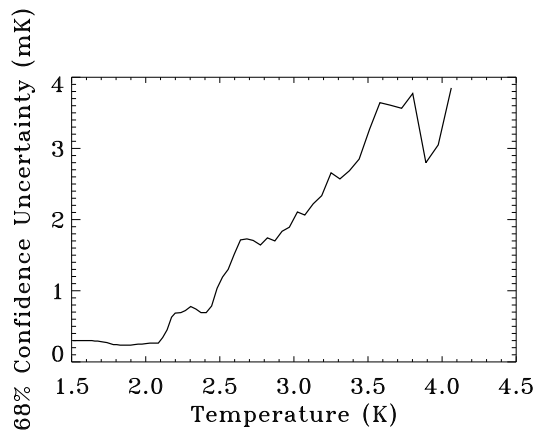}
\caption{Statistical uncertainty in the derived temperature calibration
based on scatter from 8 calibration runs in 1998 and 2002.
The uncertainty at temperatures above the superfluid transition
is dominated by thermal gradients in the bath.
\label{therm_err} }
\end{figure}

We calibrate the thermometers {\it in situ}
by immersing the entire calibrator
in a liquid helium bath
and monitoring the resistance of each thermometer
while slowly lowering the pressure above the bath.
A NIST-traceable reference thermometer
records the bath temperature
using the same readout as the calibrator thermometers.
Calibrations 4 years apart
using two different readout systems
agree within the statistical uncertainties ($\sim$ 0.3 mK).
The initial calibration in 1998
used a 1 $\mu$A excitation current
sinusoidally modulated at frequency 1 Hz.
A lockin circuit demodulates the voltage drop
across each thermometer
and integrates for 5.6 seconds
to yield the thermometer resistance.
The 1998 calibration sequentially
measured the 5 embedded thermometers
plus the reference bath thermometer.
Each thermometer was thus sampled once every 33.4 seconds
as the bath temperature
fell from 4.2 K to 1.4 K
over the course of 10 hours.
To prevent bath stratification
from producing systematic thermal gradients in the dewar,
we use only data taken
when the bath temperature was falling fast enough
to ensure vigorous boiling.
We assume that the bath and calibrator are isothermal
and interpolate the measured bath temperature
to the time recorded for each calibrator resistance measurement.

We calibrated the thermometers a second time in 2002
using the flight thermometer readout system
\cite{fixsen/etal:2002}.
The 2002 calibration differed from the 1998 calibration
in several important aspects.
The 2002 calibration used a faster square-wave modulation at 75 Hz,
applying power to each thermometer for only 26.7 ms
to reduce self-heating,
and sampling each thermometer once every 1.067 s.
A set of 4 reference resistors monitored any drifts in the readout.
Since we used the the same system
for ground calibration as flight,
effects such as 
self-heating or
the shunt impedance from
stray capacitance in the leads
are automatically accounted for.
The 2002 calibration
also included the two control thermometers
mounted on the copper heater plate.

We obtained 3 calibration runs in 1998
and 5 more in 2002.
Direct comparison between years
requires a correction for self-heating.
Although self-heating in the 2002 calibration
is the same as in flight,
it is not the same as the 1998 calibration
which applied power to each thermometer
for a much longer duration.
Additional tests in 1998
measured the change in thermometer temperature
as the excitation current
varied from 1 $\mu$A to 10 $\mu$A
at a series of bath temperatures ranging from 1.6 K to 4.2 K.
The resulting self-heating is well described by 
\begin{equation}
\Delta T = \Delta T_0 
~\left( \frac{I}{1 ~\mu {\rm A}} \right)^2
~\left( \frac{T}{1 ~{\rm K}} \right)^{-2.2}
\label{slef_heat_eq}
\end{equation}
where
$\Delta T_0 \sim 27$ mK (varying slightly among the thermometers)
and the fitted power law in temperature
includes the dependence
of both the thermometer resistance 
and the thermal conductance of the absorber.
We correct the 1998 data for self-heating
but do not explicitly correct the 2002 data
since the effect is included in both the ground and flight readout.

We interpolate the measured resistance data
to generate a resistance to temperature calibration
for each thermometer
over the temperature range 1.4 to 4.2 K.
The calibrations derived from each of the 8 individual runs
are in good agreement with each other.
We quantify this
by generating a set of ``standard'' resistance values
spanning the observed range,
then converting these to the equivalent temperature
using the resistance-temperature calibration curves
from each of the 8 separate calibration runs.
The resulting scatter in the computed temperatures
at each resistance
shows the agreement from run to run.
Figure \ref{therm_err} shows the 68\% confidence uncertainty
in the mean for thermometer T1
as a function of bath temperature.
Below the superfluid transition
all calibrations agree within 0.3 mK,
consistent with the noise
of the thermometer readout
\cite{fixsen/etal:2002}.
Above the transition temperature
the uncertainty increases linearly with the bath temperature.
This excess noise is not observed at colder temperatures,
suggesting an origin from thermal fluctuations
within the liquid helium bath itself.

\begin{figure}[t]
\includegraphics[width=3.25in]{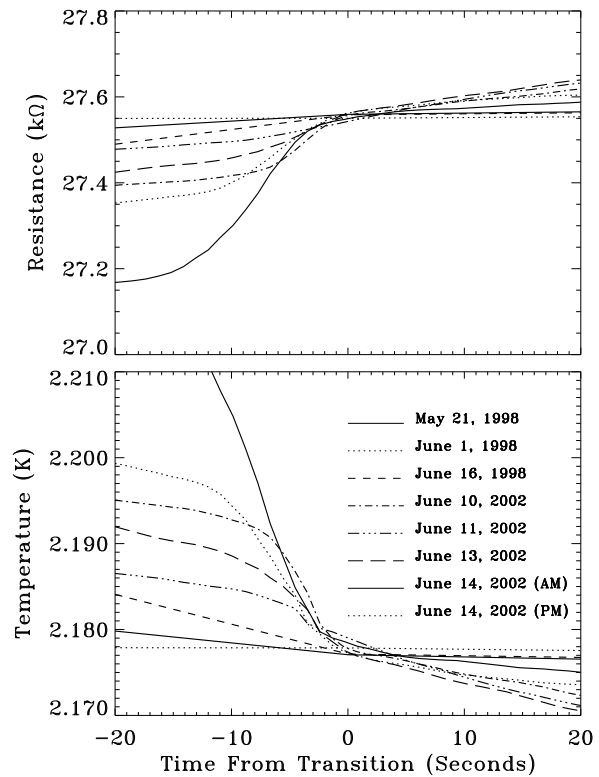}
\caption{Resistance and calibrated temperatures for calibrator thermometer T1
from all calibration runs, as the bath temperature passes through
the superfluid transition.  
The superfluid transition creates a distinctive knee in 
the time-ordered data.
(top) Measured resistance values.
(bottom) Calibrated temperatures.
The calibrated temperature for each run
correctly reproduces the transition at T=2.1768 K.
\label{lambda_vs_time} }
\end{figure}

We use the mean of the 8 individual calibration runs
to produce a single resistance--temperature calibration
for each thermometer.
Observations of the superfluid helium transition 
at temperature 2.1768 K
provide an independent check on the absolute calibration.
We record the resistance of each thermometer during each run
as the pressure above the bath slowly falls.
The superfluid transition 
produces a distinctive kink
in the time-ordered data
as the thermal properties of the bath abruptly change.
Figure \ref{lambda_vs_time} shows all observations 
of the superfluid transition
for thermometer T1 from the 1998 and 2002 calibration runs.
The calibrated temperatures correctly reproduce
the known transition temperature.

Additional data in 2003 provide thermometer calibration
at temperatures from 4.2 to 20 K.
This higher temperature calibration
placed the calibrator inside a copper cryostat
mounted within a liquid helium bucket dewar.
We used the flight thermometer readout system to measure
the resistance of each flight thermometer
as the liquid level in the dewar fell below the calibrator enclosure.
The absorber temperature rose from 4.2 K to 20 K
over a period of 20 hours,
much longer than the few second time constant of the calibrator,
ensuring near isothermality across the absorber.
Data taken with the absorber immersed in liquid helium
agreed with previous measurements from the 1998 and 2002 calibrations.

We conclude that the ARCADE external calibrator
meets the requirements
for measurements of the CMB blackbody spectrum.
The calibrator achieves emissivity greater than 0.9997
within a compact footprint.
The calibrator thermometry
is stable in time over 4 years,
with statistical uncertainty in the temperature calibration
of order 2 mK near 2.7 K,
limited primarily by thermal fluctuations in the liquid helium bath.
Observations of the superfluid transition
demonstrate that the absolute temperature scale
is accurate within 0.3 mK.

\begin{acknowledgments}
We thank M. DiPirro and D. McHugh
of the Cryogenic Fluids Branch at GSFC
for supporting the ARCADE thermal calibration.
This material is based on work supported by
the National Aeronautics and Space Administration
under Space Astrophysics and Research Analysis
program of the Office of Space Science.
\end{acknowledgments}

\bibliography{target_cal_astro}

\end{document}